\documentclass[preprint]{aastex}

\usepackage{natbib}

\newcommand{\lya}{Ly$\alpha$ }

\shorttitle{A Projected Pair of QSOs}
\shortauthors{Heintz et al.}

\begin{document}


\title{Serendipitous discovery of a projected pair of QSOs separated by 4.5
arcsec on the sky} 


\author{K.~E.~Heintz\altaffilmark{1} and J.~P.~U.~Fynbo\altaffilmark{1} and J.-K.~Krogager\altaffilmark{1,7} and M.~Vestergaard\altaffilmark{1,2}}
\affil{Dark Cosmology Centre, Niels Bohr Institute, University of Copenhagen, Juliane
Maries Vej 30, 2100 Copenhagen O, Denmark
and
Steward Observatory, University of Arizona, 933 N. Cherry Avenue, Tucson AZ 85721, USA}
\email{heintz@dark-cosmology.dk}

\author{P.~M\o ller\altaffilmark{3} and M.~Arabsalmani\altaffilmark{3}}
\affil{European Southern Observatory, Karl-Schwarzschildstrasse 2, D-85748 Garching, Germany}

\author{S.~Geier\altaffilmark{4,5}}
\affil{Gran Telescopio Canarias (GRANTECAN), Cuesta de San Jos\'e s/n, E-38712 , Bre\~na Baja, La Palma, Spain 
and
Instituto de Astrof\'isica de Canarias, V\'ia L\'actea s/n, E38200, La Laguna, Tenerife, Spain}

\author{P.~Noterdaeme\altaffilmark{6}}
\affil{Institut d'Astrophysique de Paris, CNRS-UPMC, UMR7095, 98bis bd Arago, F-75014 Paris, France}

\author{C.~Ledoux\altaffilmark{7}}
\affil{European Southern Observatory, Alonso de C\'ordova 3107, Vitacura, Casilla 19001, Santiago 19, Chile}

\author{F.~G.~Saturni\altaffilmark{8}}
\affil{University of Rome "La Sapienza", p.le A. Moro 5, I-00185 Rome, Italy}

\and

\author{B.~Venemans\altaffilmark{9}}
\affil{Max-Planck Institute for Astronomy, K{\"o}nigstuhl 17, D-69117 Heidelberg, Germany}


\altaffiltext{1}{Based on observations made with the Nordic Optical
Telescope, on the island of La Palma jointly operated by Denmark, Finland,
Iceland, Norway, and Sweden, in the Spanish Observatorio del Roque de los
Muchachos of the Instituto de Astrofisica de Canarias.}


\begin{abstract}
We present the serendipitous discovery of a projected pair of quasi-stellar objects (QSOs) with an
angular separation of $\Delta\theta =4.50$ arcsec. The redshifts of the
two QSOs are widely different: one, our programme target, is a QSO
with a spectrum consistent with being a narrow line Seyfert 1 AGN
at $z=2.05$. For this 
target we detect Lyman-$\alpha$, \ion{C}{4}, and \ion{C}{3]}. 
The other QSO, which
by chance was included on the spectroscopic slit, is a Type 1 QSO at
a redshift of $z=1.68$, for which we detect  \ion{C}{4}, \ion{C}{3]} and
\ion{Mg}{2}. We compare this system to previously detected
projected QSO pairs and find that only about a dozen previously known pairs have
smaller angular separation.
\end{abstract}





\section{Introduction} \label{sec:introduction}

We report the discovery of a closely projected pair of quasi-stellar objects (QSOs) with an angular separation of only $\Delta\theta=4.50$ arcsec. The
observing run from which the spectra of these two objects were obtained was
unrelated to the search for close projected QSO pairs. Originally, the run was
designed to spectroscopically verify candidate dust-reddened QSOs, as part of
the \textit{High A$_V$ Quasar (HAQ)} survey \citep{Fynbo13,Krogager15}.
The dust-reddened candidate QSO originally intended to be
verified spectroscopically, \textit{HAQ2358+1030A}, J2000 coordinates
($\alpha_A=23^\mathrm{h}~58^\mathrm{m}~40.47^\mathrm{s},~\delta_A=+10^\mathrm{o}~30^\mathrm{m}~40.07^\mathrm{s}$) will henceforth just be denoted
as object A. The companion \textit{"HAQ"2358+1030B} with J2000
coordinates ($\alpha_B=23^\mathrm{h}~58^\mathrm{m}~40.53^\mathrm{s},~\delta_B=+10^\mathrm{o}~30^\mathrm{m}~35.53^\mathrm{s}$) will be denoted as
object B. The companion QSO was discovered serendipitously, coincidentally
placed on the slit during the target acquisition of object A. The
slit was aligned at the parallactic angle.

The small angular separation of this projected pair of QSOs is quite unusual
and will be discussed briefly in Sec.~\ref{sec:conc}. Previously,
\citet{Hennawi06a} have carried out an extensive search for binary QSO systems, using
the Sloan Digital Sky Survey \citep[SDSS;][]{York00} and the 2dF QSO Redshift Survey \citep[2QZ;][]{Croom04} QSO catalogs. They primarily
focused on selecting binary QSO pairs, the controversial population discovered in the search for small scale $2\le \Delta\theta \le 10$ arcsec gravitationally lensed QSOs \citep[e.g.,][]{Mortlock99}, to study the small-scale QSO
clustering and correlation function. However, the majority of the detected QSO pair systems were
projected systems, having relative radial separations of $\Delta
z\approx0.3-1.0$. The discovery of these close angularly separated projected QSO pairs initiated the search and study of "quasars probing quasars" \citep[see e.g.,][and later papers]{Hennawi06b}.
Close binary QSOs and projected QSO pairs
are important for studies of: small-scale QSO clustering, the tomography of
the inter-galactic medium along close line of sights, effects of QSO transverse ionization \citep{Møller92} and gravitationally lensed
QSOs.

Throughout the paper we assume the standard $\Lambda CDM$ cosmology with $H_0=70$ km s$^{-1}$
Mpc$^{-1}$, $\Omega_M=0.28$ and $\Omega_{\Lambda}=0.72$ \citep{Komatsu11}.

\section{Observations} \label{sec:data}

\begin{figure}
\epsscale{.80}
\plotone{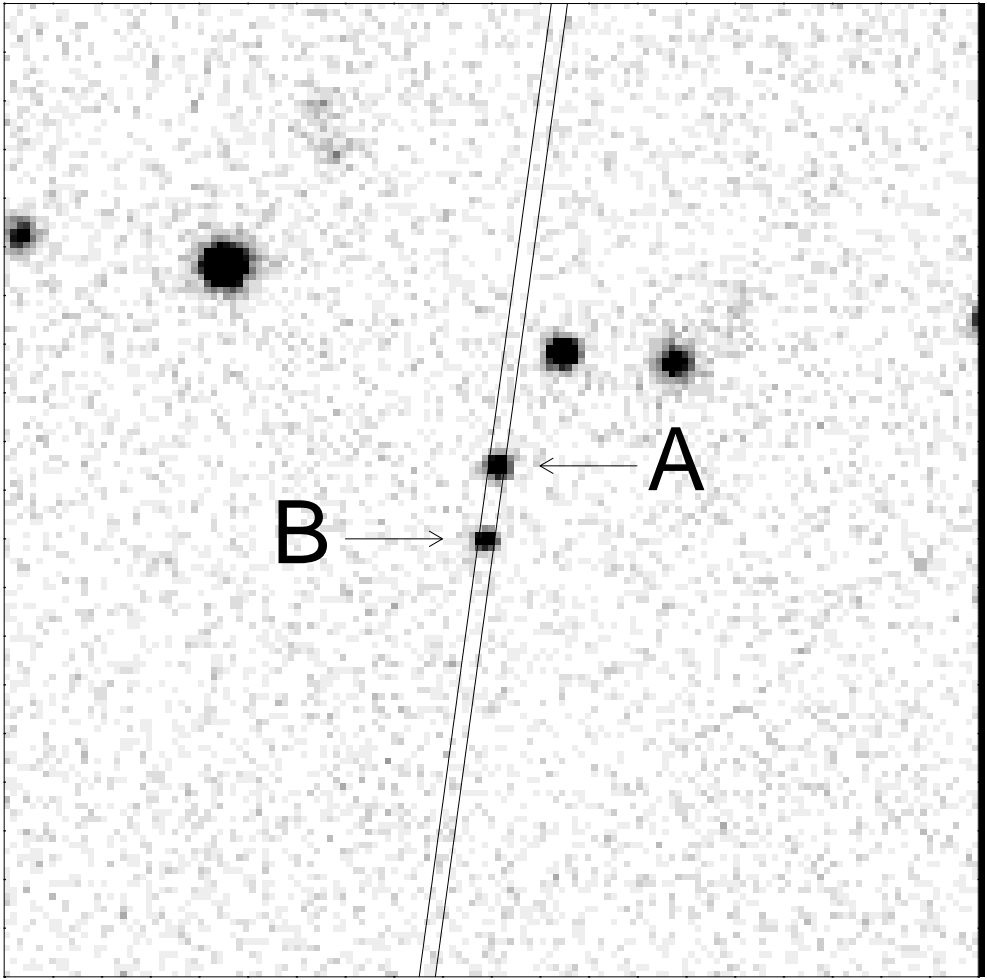}
\caption{A 1$\times$1 arcmin$^2$ field around the two sources
(marked A and B) as imaged in the $i$-band by SDSS DR12. North is up and East is to the left. A schematic view of the actual slit, used during the observation, which was centred on source A and aligned with the parallactic angle
(-7.7$^\mathrm{o}$ East of North (EoN) at the time) is shown. The position angle between the
two objects is -10.1$^\mathrm{o}$ EoN.}
\label{fig:findchart}
\end{figure}

In Fig.~\ref{fig:findchart} we present a 1$\times$1 arcmin$^2$ field around the two sources
(marked A and B) as imaged in the $i$-band by SDSS in DR12 \citep{Alam15}.

The observation was carried out during an observing run with the Nordic Optical
Telescope (NOT) on La Palma in September, 2015. The spectra were obtained using
the Andalucia Faint Object Spectrograph and Camera (ALFOSC), grism \#4 covering the wavelengths $3200-9100$~\AA~(with a spectral resolution of 21~\AA) and a
slit width of 1.3 arcsec. Blocking filter \#94 was used in
order to prevent second-order contamination from wavelengths shorter than 3560
\AA. Two exposures of 1200s were taken. The object south of our programme target is the one captured serendipitously.

One additonal spectrum was taken on January 9 2016 in a redder grism 
(grism \#20). This spectrum
was taken again with a position angle covering both QSOs. A total exposure
time of 3200 sec was secured. The purpose of this spectrum was to confirm
that an unidentified emission line in the spectrum of object A was due 
to second order contamination (see below).

The spectra were processed using a combination of IRAF\footnote{IRAF is the Image
Reduction and Analysis Facility, a general purpose software system for the
reduction and analysis of astronomical data. IRAF is written and supported by
the National Optical Astronomy Observatories (NOAO) in Tucson, Arizona. NOAO is
operated by the Association of Universities for Research in Astronomy (AURA),
Inc. under cooperative agreement with the National Science Foundation} and
MIDAS\footnote{ESO-MIDAS is a copyright protected software product of the
European Southern Observatory. The software is available under the GNU General
Public License.} tasks for low resolution spectroscopy. To reject cosmic rays
we used the La\_cosmic \citep{vanDokkum01}.
We corrected the spectra for Galactic extinction using the
extinction maps of \cite{Schlegel98}. To improve the absolute flux
calibration we scaled the spectra to be consistent with the r-band 
photometry from SDSS.




\section{The projected QSO pair HAQ2358+1030A and B}\label{sec:results}

The projected angular separation between the two objects is 
$\sim4.6~\mathrm{arcsec}$ as measured on the SDSS images\footnote{http://skyserver.sdss.org/dr12/en/tools/explore/Summary.aspx?}.
Based on the acquisition picture obtained we
find a projected angular separation of only
$23.5~\mathrm{pixels}=4.50~\mathrm{arcsec}$ (0.19 arcsec/pixel for ALFOSC). This corresponds to a physical angular separation of 39.2 kpc at $z=1.5$.
Since our measurement is done directly from the acquisition picture, we consider this a more precise estimate of the angular separation of the two objects due to the better spatial resolution in these data.

\begin{figure}[ht]
\epsscale{.90}
\plotone{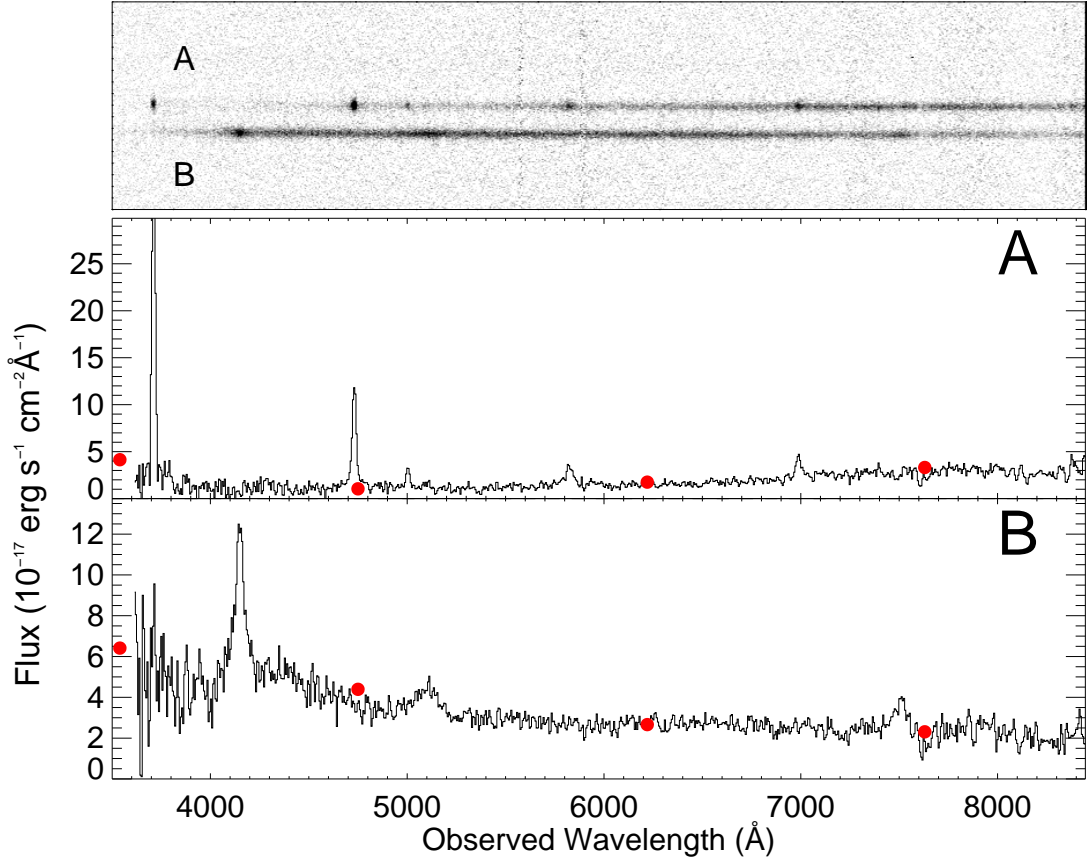}
\caption{Top panel: The 2D spectrum showing the traces of the two sources marked with A and B. Middle and bottom panels: The 1D spectra of the two sources.
The observed spectra are plotted as the solid black lines and the photometric data points from SDSS are shown as the red dots (left to right: $u,~g,~r~,i$). The spectra have been scaled to match the r-band photometric data point from SDSS.
Object A is a QSO equivalent of a narrow line Seyfert 1 galaxy at $z=2.05$ whereas B is a Type 1 QSO at $z=1.68$.}
\label{fig:Q2358+1030ab}
\end{figure}

Fig.~\ref{fig:Q2358+1030ab} shows the two one-dimensional spectra after
flat-field correction, bias and sky subtraction and flux calibration along with
the photometry from SDSS and the UKIRT Infrared Deep Sky Survey \citep[UKIDSS;][]{Lawrence07} which are all on the AB magnitude system. We
determine the redshift by visual inspection of the emission lines visible in
the spectra. Table~\ref{tab:mag} lists the optical and near-infrared AB magnitudes for each of the two objects from the SDSS DR12 and the UKIDSS DR10plus catalogs.

\begin{table*}[!htbp]
    \caption{The optical and near-infrared magnitudes of object A and B all on the AB magnitude system from the SDSS and UKIDSS catalogs.}
\begin{tabular*}{1.0\textwidth}{@{\extracolsep{\fill}}l c c c c c c c c c }
    \noalign{\smallskip} \hline \hline \noalign{\smallskip}
        \emph{Object} & \emph{u} & \emph{g} & \emph{r} & \emph{i} & \emph{z} & \emph{Y} & \emph{J} & \emph{H} & \emph{K$_s$}\\ 
          & mag & mag & mag & mag & mag & mag & mag & mag & mag
         \\
        \hline
        A & 21.37 & 22.10 & 20.82 & 19.62 & 18.92 & 18.62 & 18.21 & 17.71 & 17.24 \\
        
        B & 20.90 & 20.54 & 20.37 & 20.01 & 20.01 & 20.06 & 20.07 & 19.74 & 19.71 \\
        \noalign{\smallskip} \hline \noalign{\smallskip}
    \end{tabular*}
    \label{tab:mag}
    \end{table*}

In Table~\ref{tab:ABdata} we list the wavelengths, line widths (full width at
half maximum, FWHM), and derived redshifts for the detected emission lines
for the two objects. Object A is identified to be a dust reddened
QSO equivalent of a narrow line Seyfert 1 galaxy classified by its narrow
\ion{C}{4}$~\lambda 1549$ emission line with a FWHM of $920~\mathrm{km/s}$.  The redshift is determined to be
$z=2.053$ based on the visible \lya and the \ion{C}{4}, \ion{C}{3]}$~\lambda 1909$ emission lines. There is an additional emission line
in this spectrum at rest wavelength 2289 \AA. This line is not prominent in the
the empirical ultraviolet template for iron emission in AGN derived 
from I Zwicky 1 \citep{Vestergaard01} and we have not been able to find it
in other spectra for QSOs in the literature. 
The new spectrum obtained in Januar 2016 at the NOT with grism \#20 confirms that this line is due to second order contamination \citep{Stanishev07} and is hence not real.

The reddening of this object was determined to be $A_V=1.1$ from the photometric
data points and the shape of the continuum following the procedure of 
\citet{Fynbo13} and \citet{Krogager15}. Object B is identified to be a
regular, unreddened Type 1 QSO with a FWHM of $4200~\mathrm{km/s}$ of
the \ion{C}{4} emission line. The redshift was determined to be
$z=1.68$ based on the broad \ion{C}{4} and \ion{C}{3]}
emission lines. The unobscured nature of this object was again determined from
the photometric data points and the shape of the continuum.

The redshift measurements infer a relative radial distance of $\Delta z = 0.38$
between the two QSOs. Only a dozen other systems of projected QSO pairs with
$\Delta\theta<4.50~\mathrm{arcsec}$ have been reported in the systematic search
by \citet[see tables 8 and 9]{Hennawi06a}, \cite{Inada12} and \cite{More16}, all with
relative radial distances $\Delta z \sim 0.3-1.1$. It is also worth noting that
neither of these two QSOs are included in the DR12 QSO (DR12Q)
survey. We selected object A as a candidate dust reddened QSO in the HAQ survey \citep[see
e.g.,][]{Fynbo13,Krogager15}. Object B evaded selection due to its specific
photometry falling outside the selection criteria of BOSS, which is optimized
for $z>2$.

\begin{figure} 
\epsscale{.90}
\plotone{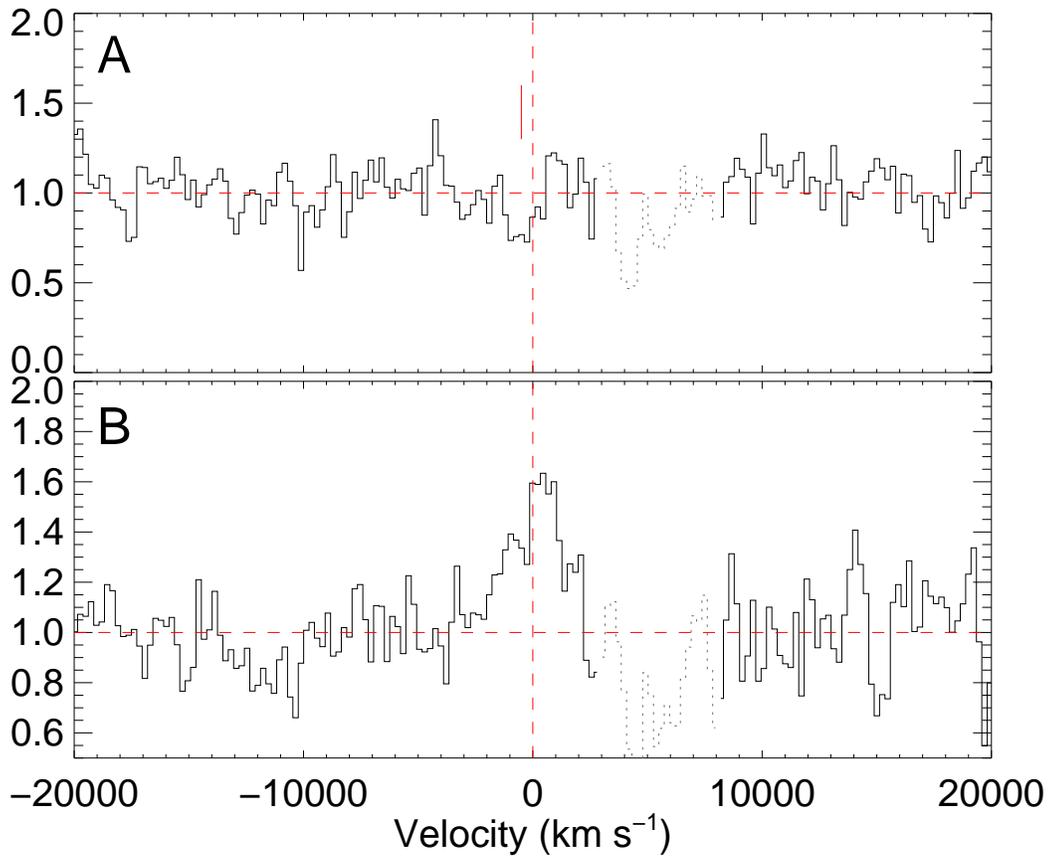}
\caption{This figure shows the normalized spectra of object A and B in the
regions around \ion{Mg}{2} at the redshift of object B. In the top panel there
is a hint of \ion{Mg}{2} absorption with a formal significance of about
3$\sigma$. The part of the spectra shown in dotted grey represents telluric
absorption.}
\label{fig:abs}
\end{figure}

We looked for associated absorption in the two spectra, more specifically
absorption in the spectrum of object A due to, e.g., \ion{Mg}{2}$~\lambda 2800$ or
\ion{C}{4} due to gas associated with the foreground QSO object B. The
continuum level in the spectrum of A is too weak at the position of \ion{C}{4} at the redshift of object B
to allow detection of absorption. For \ion{Mg}{2} we show the relevant spectral regions in
Fig.~\ref{fig:abs}. There is a hint of \ion{Mg}{2} absorption at 7490.2 \AA \
with an observed equivalent width of 8$\pm$3 \AA \ found by integrating
the error spectrum across the aborption line extent. This wavelength corresponds
to $z_\mathrm{abs} = 1.676$, roughly consistent, but slightly blueshifted, 
relative to the expected position at the redshift of object B.

\section{The significance of closely separated QSO pairs} \label{sec:conc}

The projected pair of close angularly separated QSOs is a rare finding.
\cite{Richards05} find an integrated space density of QSOs brighter
than $g=20.5$ (brightness of object B) of  $\sim 31~\mathrm{deg}^{-2}$
based on the 2dF-SDSS LRG and QSO (2SLAQ) survey. This intrinsic number
density is based on robust optical five-band ($u,g,r,i,z$) color-color cuts
similar to those utilized in the first SDSS QSO selection algorithm
\citep{Richards02} although with slight modifications for the faintest objects.
Based on the integrated space density, we estimated the likelihood of observing 
one additional QSO in the slit placed on object A. Considering only QSO pairs where (as in our case) the redshift
difference corresponds to a distance much larger than the
2-point correlation length, then the likelihood can be calculated as:
\begin{equation}
\sim(31/3600/3600)\times 1.3 \times 10 \approx 3.1\times 10^{-5},
\end{equation}
where the last two numbers are the width of the slit and the length within which
object B was detected ($\sim 2\times 5$ arcsec) from object A, respectively. Had the observation 
been carried out just an hour later, the parallactic angle of the slit would not 
have captured this nearby companion, and we would not have
discovered this system.
When extrapolating the conservative total number of QSOs in the sky of 
$\sim 31\times 41.253~\mathrm{deg}^{2} \approx 1.3\times 10^6$ and using the estimate of the 
likelihood per square degree we compute how many such cases on the entire sky is expected. 
We found a total of only $\sim 140$ cases of such closely separated QSO systems within the 
magnitude limit of object B. This detection was indeed serendipitous.

\begin{table*}[!htbp]
\caption{Wavelength, line widths (measured and corrected for the spectral resolution) 
and derived redshifts for all detected emission lines.}
\begin{tabular*}{1.0\textwidth}{@{\extracolsep{\fill}}l c c c c }
\noalign{\smallskip} \hline \hline \noalign{\smallskip}
 \emph{Line} & \emph{$\lambda_{obs}$} & FWHM  &  FWHM$_{corr}$ & $z$ \\
             &      {\AA}             &{\AA}  & {\AA}          &    \\
\hline
Object A &              & &&\\
\hline
Ly$\alpha$ $\lambda$ 1216    & 3711.8$\pm$0.5  &  20.0$\pm$1.2 & 12.3$\pm$1.2 & 2.0533$\pm$0.0004 \\
\ion{C}{4} $\lambda$ 1549  & 4731.0$\pm$0.4  &  24.8$\pm$0.9 & 14.5$\pm$0.9 & 2.0533$\pm$0.0004 \\
\ion{He}{2} $\lambda$ 1640 & 5003.7$\pm$1.3  &  20.0$\pm$0.9 & -- & 2.0503$\pm$0.0008 \\
\ion{C}{3]} $\lambda$ 1909 & 5825.1$\pm$1.8  &  43.1$\pm$4.5 & 35.3$\pm$4.5 & 2.0530$\pm$0.0008 \\
Ly$\alpha$ 2nd order $\lambda$ 2289  & 6989.7$\pm$1.4  &  36.2$\pm$3.4 & 20.7$\pm$3.4 & 2.053 \\
\hline
Object B &              & &&\\
\hline
\ion{C}{4} $\lambda$ 1549  & 4149.7$\pm$1.7  &  61.0$\pm$4.1   & 58.4$\pm$4.1   & 1.6781$\pm$0.0011 \\
\ion{C}{3]} $\lambda$ 1909 & 5100.8$\pm$5.8  &  124.4$\pm$14.4 & 122.5$\pm$14.4 & 1.673$\pm$0.003 \\
\ion{Mg}{2} $\lambda$ 2800 & 7506.4$\pm$3.8  &  63.1$\pm$10.1  & 54.5$\pm$10.1  & 1.6817$\pm$0.0014 \\
\noalign{\smallskip} \hline \noalign{\smallskip}
\end{tabular*}
\label{tab:ABdata}
\end{table*}




\newpage

\acknowledgments
We wish to thank the anonymous referee for the helpful comments improving the
quality of this work.
The research leading to these results has received funding from the European
Research Council under the European Union's Seventh Framework Program
(FP7/2007-2013)/ERC Grant agreement no. EGGS-278202.
The data presented here were obtained with ALFOSC, which is provided by the Instituto de Astrofisica de Andalucia (IAA) under a joint agreement with the University of Copenhagen and NOTSA. BV acknowledges funding through the ERC grant “Cosmic Dawn”.
MV gratefully acknowledge support from the Danish Council for Independent Research via grant no. DFF 4002-00275.
Funding for the Sloan Digital Sky Survey III has been provided by
the Alfred P. Sloan Foundation, the U.S. Department of Energy Office of
Science, and the Participating Institutions. SDSS-IV acknowledges
support and resources from the Center for High-Performance Computing at
the University of Utah. The SDSS web site is www.sdss.org.

\clearpage

\end{document}